\documentclass[preprint2]{emulateapj}
\usepackage{color}

\slugcomment{published in ApJ 712 (2010) L16, Received 2010 January 21; accepted 2010 February 11}
\shorttitle{Electronic Spectroscopy of PAHs}
\shortauthors{Steglich et al.}

\begin{document}
\title{Electronic Spectroscopy of medium-sized Polycyclic Aromatic Hydrocarbons: Implications for the Carriers of the 2175 \AA\ UV Bump}
\author{M. Steglich\altaffilmark{1}, C. J\"ager, G. Rouill\'e, F. Huisken}
\affil{Laboratory Astrophysics Group of the Max Planck Institute for Astronomy at the Friedrich Schiller
University Jena,\\
Institute of Solid State Physics, Helmholtzweg 3, D-07743 Jena, Germany}
\email{Mathias.Steglich@uni-jena.de}
\email{Cornelia.Jaeger@uni-jena.de}
\author{H. Mutschke}
\affil{Astrophysical Institute and University Observatory, Friedrich Schiller University Jena, Schillerg\"asschen 3, D-07745 Jena, Germany}
\author{Th. Henning}
\affil{Max Planck Institute for Astronomy, K\"onigstuhl 17, D-69117 Heidelberg, Germany}
\altaffiltext{1}{Author to whom any correspondence should be adressed.}

\begin{abstract}
Mixtures of polycyclic aromatic hydrocarbons (PAHs) have been produced by means of laser pyrolysis. The main fraction of the extracted PAHs was primarily medium-sized, up to a maximum size of 38 carbon atoms per molecule. The use of different extraction solvents and subsequent chromatographic fractionation provided mixtures of different size distributions. UV--VIS absorption spectra have been measured at low temperature by matrix isolation spectroscopy and at room temperature with PAHs as film-like deposits on transparent substrates. In accordance with semi-empirical calculations, our findings suggest that large PAHs with sizes around 50--60 carbon atoms per molecule could be responsible for the interstellar UV bump at 217.5 nm.
\end{abstract}


\section{INTRODUCTION}
Primary cosmic carbonaceous matter is produced in envelopes of asymptotic giant branch stars via gas-phase condensation \citep{Henning98}. Important components are large dust particles and molecules, such as polycyclic aromatic hydrocarbons (PAHs). Another source of PAHs can be the shattering of carbonaceous grains in interstellar shocks \citep{Tielens08}. The existence of PAHs in several astrophysical environments, ranging from planetary nebula, reflection nebulosities, circumstellar disks to even active galactic nuclei, has been inferred from the presence of infrared emission bands related to C--H and C--C vibrational modes \citep{Leger84}. However, the identification of a specific molecule based on these aromatic bands in the mid-infrared is not possible. Only information about the size and charge state distributions, as well as the ratio between aliphatic and aromatic carbons can be obtained \citep{draine07}. Based on energetic arguments, the infrared emission is attributed to aromatic molecules containing about 50 carbon atoms \citep{Tielens08}. After H$_2$ and CO, PAHs are probably the most abundant molecules in the interstellar medium \citep[ISM;][]{Leger89}. It is possible that they are also responsible for a few other, still unexplained features like the extended red emission \citep[ERE;][]{Rhee07}, the blue luminescence \citep{Vijh05}, the extinction bump at 217.5 nm \citep{beegle97}, and especially the diffuse interstellar bands \citep[DIBs;][]{Jenniskens94, Salama99, ruiterkamp02}. For recent reviews see, e.g., \citet{Tielens08} and \citet{salama08}.

Theoretical and experimental work has been focused on wavelengths longer than 200 nm (5 $\mu$m$^{-1}$). Only a few studies dealt with the far-UV behavior of PAHs. \citet{joblin92} measured the absorption between 1 and 11 $\mu$m$^{-1}$ of hot (600 K), gaseous PAHs, containing about 30 carbon atoms, and they proposed that these PAHs may give a strong contribution to the overall interstellar extinction as well as to the UV bump.

Nano-sized amorphous carbon structures have been often considered to explain the UV bump \citep{papoular96, menella98, schnaiter98}. Possible carriers of the bump also include naphthalene-based aggregates \citep{arnoult00, beegle97} or strongly dehydrogenated PAHs \citep{duley98, duley04, malloci08}. Ab initio calculations carried out by \citet{malloci04} and \citet{pestellini08} have shown that a mixture of PAHs in different charge states could also give rise to a bump around 217.5 nm.

In this Letter, we report on UV--VIS absorption spectroscopy of PAH mixtures that were produced by laser pyrolysis with subsequent size separation. Spectra were measured for films of PAHs at room temperature and for molecules isolated in neon at 6 K. We have compared the experimental results with those of semi-empirical calculations of mixtures containing different size distributions of PAH molecules. From our measurements, we derive conclusions about the connection between PAHs and the general interstellar extinction, including the bump  at 217.5 nm.

\section{SYNTHETIC SPECTRA OF PAH MIXTURES}
\label{zindo}
We calculated PAH electronic spectra, using, for ground state geometry optimizations, the Austin model 1 \citep[AM1;][]{dewar85}, as implemented in the Gamess-US software \citep{schmidt93}, and, for the determination of electronic excitation energies, Zerner's model of intermediate neglect of differential overlap \citep[ZINDO;][]{ridley73}, as implemented in the Gaussian software \citep{frisch04}. The ZINDO technique is known to give fairly good results for PAHs at low computational cost. \citet{ruiterkamp05} used it to determine the PAH size and charge state distribution in the ISM by comparing the calculated transition energies with the DIB spectrum.

However, we are aware of the limitations of the ZINDO technique. We consider calculated $\pi \rightarrow \pi ^{\ast}$ transitions to be reliable whereas the oscillator strengths and energy positions of transitions involving $\sigma$ electrons are less accurate. Our model covers 122 different PAHs containing between 10 and 72 carbon atoms per molecule. We included compact as well as elongated structures. Even though this procedure can never take all molecules of the ISM into account, we think that our set of PAHs gives a good indication about the energetic positions of electronic transitions and the spectral range where a high density of states can be expected. Spectra of single PAHs were determined by assuming a Lorentzian profile with the calculated oscillator strength as area and taking a line width of 5000 cm$^{-1}$. Then these spectra were weighted according to a size distribution and summed up. We have chosen such a large line width to account for vibronic patterns, which consist of close-by lying absorption bands. The spectra thus obtained cannot be compared with gas-phase measurements, but are in good agreement with measured spectra of solid PAH deposits (see Section \ref{results}).

\section{EXPERIMENTAL DETAILS}
\label{experiment}
Details about the setup for laser pyrolysis and the connection to particle condensation in astrophysical environments can be found in a previous paper \citep{Jaeger07}. Briefly, a 60 W continuous wave CO$_2$ laser decomposes gaseous hydrocarbons. Molecules and particles are created in the hot (1100$\degr$C), flame-like condensation zone and can be collected afterward. In this study, soluble molecules (PAHs) were extracted from the condensate firstly with methanol and secondly with dichloromethane (DCM). Compositions of the various extracts were characterized by high-performance liquid chromatography (HPLC), gas chromatographic/mass spectrometric (GC/MS) analysis, and matrix-assisted laser desorption/ionization in combination with time-of-flight mass spectrometry (MALDI--TOF). Results and technical details about the analytical and spectral analyses were provided previously \citep{Jaeger06,Jaeger07,Jaeger09}.

We used matrix isolation spectroscopy (MIS) to obtain low-temperature spectra of PAHs \citep[see, e.g.,][]{Bondybey96}. As matrix material, we employed 6 K cold neon (Linde, purity 99.995 \%). Transmission spectroscopy down to 190 nm was performed with a spectrophotometer (JASCO V-670 EX). To vaporize the solid PAH samples, we used two different methods. An oven can heat up the material to a maximum temperature of 400$\degr$C. Unfortunately, when working with PAH mixtures, the use of thermal vaporization can be a problem. The individual molecules have different vapor pressures, leading to a different distribution in the matrix compared to the original sample. Larger PAHs may not evaporate at all. To overcome these problems, we also used laser vaporization (second harmonic of a pulsed Nd:YAG laser) to bring the molecules into the gas phase. The power was chosen low enough to avoid fragmentation. This has been verified by HPLC analysis of the evaporated material. To extend absorption measurements down to 125 nm, we also produced films of PAHs in the same setup on CaF$_2$ windows cooled to 6 K, but without simultaneous deposition of neon. Spectra were measured in a vacuum-UV spectrophotometer (Laserzentrum Hannover).

\section{RESULTS}
\label{results}

\begin{figure}[htp]
\epsscale{1.0} \plotone{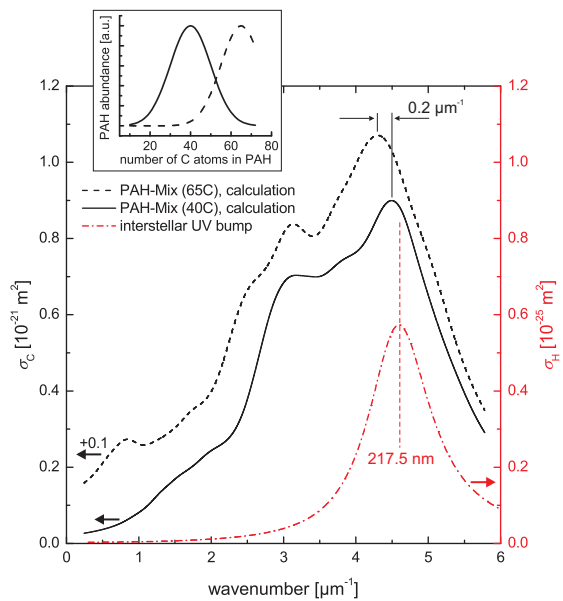} \caption{Spectra of PAH mixtures calculated with the ZINDO technique and compared with the UV bump \citep[galactic average, adapted from][]{fitzpatrick88} displayed in photoabsorption cross section per carbon atom, $\sigma _\mathrm{C}$, or hydrogen atom $\sigma _\mathrm{H}$, respectively. Inset: PAH size distributions for both synthetic spectra.} \label{fig1}
\end{figure}

Figure \ref{fig1} displays the synthetic spectra mentioned in Section \ref{zindo}. The weighted absorption curves for two different PAH size distributions centered at 40 and 65 carbon atoms are shown. The strong extinction near the position of the 217.5 nm (4.6 $\mu$m$^{-1}$) UV bump is obvious. An exact match cannot be expected at this level of theory. As already mentioned, electronic transitions involving $\sigma$ electrons may be represented too poorly by the ZINDO model. So, while the UV bump around 4.5 $\mu$m$^{-1}$ caused by $\pi \rightarrow \pi ^{\ast}$ transitions can be trusted, the calculated far-UV rise is probably too weak. Interestingly, increasing the mean size of the molecules by going from 40 to 65 carbon atoms shifts the UV bump position by roughly 10 nm (0.2 $\mu$m$^{-1}$) to the red. The shifts in the dominant absorption bands to longer wavelengths with increasing molecular size were first commented on in an interstellar context by \citet{Platt56}. If the interstellar UV bump is associated with PAHs, its constant peak position implies a size distribution with a relatively fixed mean size for the molecules. The maximum value for absorption per carbon atom at the UV bump position is almost independent of the mean size. Determined absolute values, expressed in cross section per carbon atom, are identical with calculated values from \citet{pestellini08} and measured values derived by \citet{joblin92} (for smaller PAHs). However, they should be considered as lower limit for the cross section, which could be at the maximum twice as much.

\begin{figure}[htp]
\epsscale{1.1} \plotone{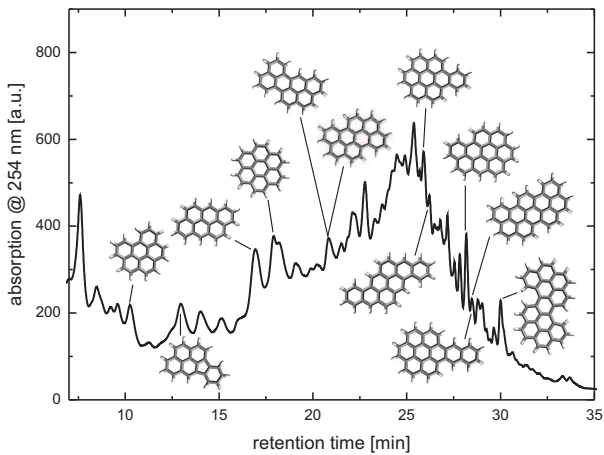} \caption{HPLC chromatogram of PAHs extracted with DCM. Identified molecules are labeled.} \label{fig2}
\end{figure}

PAH molecules produced by laser pyrolysis and extracted with DCM were analyzed by HPLC and GC/MS. All molecules identified so far (up to a maximum mass of 476 amu) can be found in \citet{Jaeger07}. An exemplary HPLC chromatogram for a certain time interval (here 7--35 minutes) is shown in Figure \ref{fig2}. Smaller PAHs reach the detector earlier and are not displayed. In the chromatogram, only the most abundant molecules or PAHs with strong extinction at 254 nm produce a clearly visible peak, all other molecules are hidden in the underlying continuum. Upon investigation of the soot extract with MALDI--TOF, it can be seen that, in principle, molecules with every integer mass (separated by 1 amu) up to 3000 amu including hydrogenated PAHs are present \citep{Jaeger09}. Small molecules up to 20 carbon atoms are far more abundant than larger ones ($>$ 30 C atoms). The fact that a high abundance of these small PAHs is incompatible with the interstellar extinction curve implies their photodestruction upon UV irradiation in the ISM.

\begin{figure}[htp]
\epsscale{1.1} \plotone{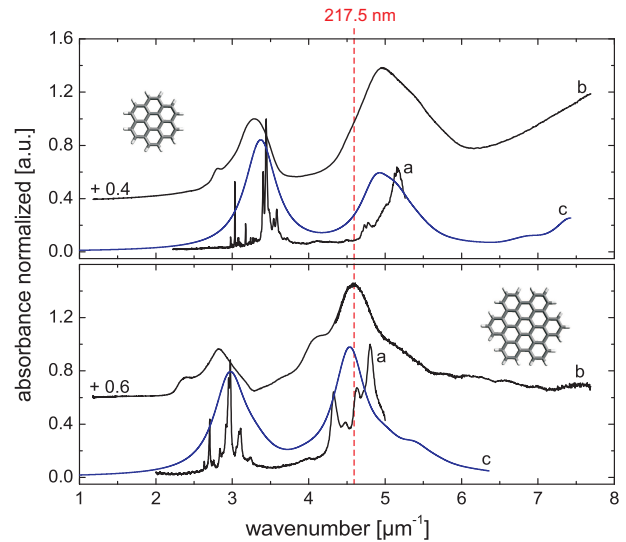} \caption{Electronic spectra of coronene (top panel) and HBC (bottom panel): matrix-isolated in neon at 6 K (a), as film on CaF$_2$ at room temperature (b), and simulated spectra using the ZINDO technique (c).} \label{fig3}
\end{figure}

Figure \ref{fig3} shows absorption spectra of coronene C$_{24}$H$_{12}$ and hexa-peri-hexabenzocoronene C$_{42}$H$_{18}$ (HBC) measured in neon matrices at 6 K. The electronic spectra of single PAHs typically consist of origin bands and associated vibronic bands that can be observed by MIS. This technique can reveal intermediate level structures caused by vibronic interactions as in the case of HBC \citep{rouille09}. Note that patterns corresponding to such structures cannot be found in the DIB spectrum. The spectrum of HBC appears shifted to the red compared to that of the smaller coronene. Both PAHs have a D$_{6h}$ symmetry and the similar patterns in the near UV range indicate that the vibronic interaction that takes place in HBC is also active in coronene. All-benzenoid PAHs to which HBC belongs are believed to be more stable against interstellar radiations \citep{troy06}. Note finally that HBC would not contribute to the far-UV rise of the interstellar extinction below 7.7 $\mu$m$^{-1}$.

Instead of being isolated, interstellar PAHs could as well be in the form of large clusters, resulting in a better protection against photodestruction. \citet{rapacioli05,rapacioli06} studied the balance between cluster formation and photoevaporation. Stacks of PAH molecules exhibit distinct spectral signatures. Due to molecular interaction, bands become broader and shift to longer wavelengths. Furthermore, depending on the size of the clusters, additional scattering has to be considered. The difference in absorption between isolated molecules and molecules interacting with each other is illustrated in Figure \ref{fig3}, where spectra of matrix-isolated coronene and HBC are compared with the corresponding spectra of film-like deposits. Scattering effects are negligible in both cases. The films have been deposited on 6 K cold CaF$_2$ windows. Upon heating to room temperature, the individual molecules of the film rearrange into positions where their potential energy is minimized. As a result, a small redshift is observed, e.g., the redshift for HBC's UV band is 3.5 nm.

For comparison, theoretical spectra calculated with the model described in Section \ref{zindo} are plotted (c). The agreement with the measured absorption spectra of PAH films is fairly good. Noteworthy, HBC films show strong UV extinction at the position of the interstellar 217.5 nm bump while the corresponding coronene bands are located further to the blue. Generally, many large PAHs absorb in the vicinity of 217.5 nm \citep[see, e.g.,][]{ruiterkamp02}. In contrast, the visible extinction depends more strongly on the molecular structure. In this way, when adding up enough spectra of single PAHs, at some point, all absorption bands will probably merge into a continuous extinction with no sharp features besides a broad bump in the UV. This idea has already been brought up by \citet{donn68}, \citet{donn69}, and later again by \citet{joblin92}.

% this figure should span both columns
\begin{figure*}[htp]
\epsscale{1.0} \plotone{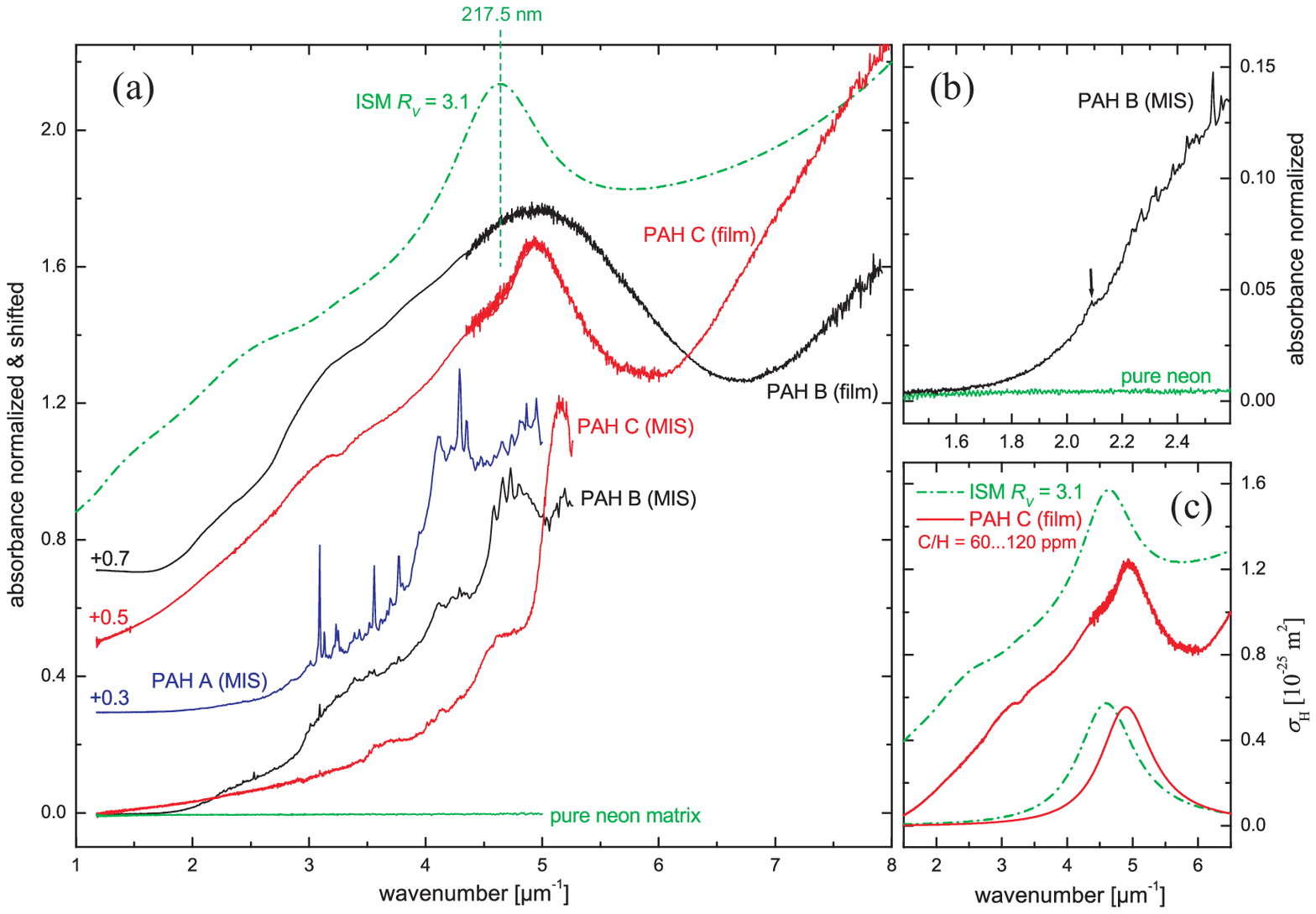} \caption{(a) MIS of different PAH mixtures. For comparison, the absorbance of a pure neon matrix of comparable thickness is shown. Furthermore, spectra of films of PAH mixtures measured in the visible, near-, and far-UV are displayed. Spectra have been shifted for clarity. Details on the samples can be found in Table 1. Experimental results are compared with the mean interstellar extinction \citep{cardelli89}. (b) Part of the spectrum ``PAH B (MIS)'' in the visible. The narrow band farthest to the red is marked by an arrow (at 2.1 $\mu$m$^{-1}$). Note that there is also absorption down to 1.6 $\mu$m$^{-1}$. (c) Mean interstellar extinction \citep{cardelli89} and experimental spectrum (PAH C), assuming $\sigma _\mathrm{C}$ = (1.5 $\pm$ 0.5)$\times$10$^{-21}$m$^2$ at the UV bump position that results in C/H = (90 $\pm$ 30)$\times$10$^{-6}$ for the shown spectrum. UV bumps have been extracted with a fitting procedure according to \citet{fitzpatrick88}.} \label{fig4}
\end{figure*}

\begin{table}[htb]
\caption{Experimental details for the samples shown in Fig. \ref{fig4}.}
\begin{tabular}{cccc}
 \hline\hline
 sample & extraction & evaporation & approx. no.\\
        &  method    &  method     & of C atoms\\
 \hline
 PAH A  & methanol$^a$      & thermal heating 400$^{\circ}$C     & 10--22  \\
 PAH B  & DCM$^b$ & thermal heating 400$^{\circ}$C     & 10--38$^d$\\
 PAH C  & 17--45 min$^c$   & Nd:YAG, 532 nm,  & 22--38$^d$\\
               &                              & 10 Hz, 6 mJ, &                    \\
               &                              & 2.5 mm spot size &                \\
 \hline
\end{tabular}

\begin{tabular}{l}
\footnotesize $^a$ methanol extract of laser pyrolysis condensate\\
\footnotesize $^b$ DCM extract of laser pyrolysis condensate\\
\footnotesize $^c$ fractionation by HPLC for a certain time window, see Fig. \ref{fig2}\\
\footnotesize $^d$ Note that larger PAHs are present (see the text).\\
\end{tabular}
\end{table}
\label{table1}

Figure \ref{fig4} displays measured electronic spectra of three different PAH mixtures. Details about the samples can be found in Table 1. The sample labeled ``PAH A'', which is essentially the methanol extract of the laser pyrolysis condensate, contains mostly small PAHs with 10--22 carbon atoms per molecule. Its low-temperature MIS spectrum is dominated by bands from pyrene C$_{16}$H$_{10}$ \citep{salama95} and phenanthrene C$_{14}$H$_{10}$ \citep{salama94} that have high vapor pressures and strong absorption bands above 3 $\mu$m$^{-1}$ ($\lambda < $ 330 nm). The strong bands are superimposed on an underlying continuous absorption. Small PAHs are still present in the DCM extract, labeled as ``PAH B'', which has been obtained from the laser pyrolysis condensate after it has been treated with methanol extraction.  For this reason, the strongest pyrene band at  3.1 $\mu$m$^{-1}$ can still be recognized. However, the main fraction of PAHs in the mixture contains molecules with 22--38 carbon atoms. It should be noted that also PAHs with masses up to 3000 amu are present in the DCM extract, but their abundance is marginal and their contribution to the spectrum can be neglected. Nevertheless, differences between ``PAH A (MIS)'' and ``PAH B (MIS)'' are clearly visible. Narrow bands start to disappear. The underlying continuous absorption develops into a single broad peak in the UV. Even though there is absorption also in the visible down to 1.65 $\mu$m$^{-1}$ (600 nm), narrow bands, similar to the DIBs, can only be found for $\lambda ^{-1} > $ 2.1 $\mu$m$^{-1}$ (480 nm). Also MIS spectra from PAH-rich coal pitch extracts showed no sharp features beyond 450 nm \citep{ehrenfreund92}. The mixture ``PAH C'' has been obtained by chromatographic fractionation of the DCM extract. It contains mainly PAHs with 22--38 carbon atoms per molecule. For this sample, laser vaporization, instead of thermal heating, has been used to prepare the matrix and also the film-like deposits. So we expect heavier molecules with low vapor pressures to contribute to the spectrum. The entire spectrum is virtually free of sharp features. Instead, strong extinction in the UV is observed with a maximum at 5.15 $\mu$m$^{-1}$ (194 nm), which means that there are enough different molecules with absorption bands at varying positions to produce a continuous extinction.

Spectra of PAH films are also displayed in Figure \ref{fig4}. Films of mixture A are not persistent under vacuum conditions due to the high vapor pressure of the small molecules. The main feature for both samples is an absorption bump at 4.93 $\mu$m$^{-1}$. The width and shape of this bump for the sample ``PAH C'' shows striking similarities with the mean interstellar extinction curve, illustrating that condensed PAHs could be the carriers for the UV bump. Considering the results of our semi-empirical ZINDO calculations, the bump position of PAH mixtures containing a distribution of molecules with a mean size of 50--60 carbon atoms will be shifted to the red, probably coming close to 4.6 $\mu$m$^{-1}$ (217.5 nm). The quite strong far-UV rise of ``PAH C'' (beyond 6 $\mu$m$^{-1}$) is somehow peculiar, but note that certain galactic lines of sight exhibit strong far-UV extinction as well \citep{fitzpatrick88}. However, calculations \citep{malloci07} and measurements (HBC in Fig. \ref{fig3}) predict a moderate or rather weak slope in the far-UV extinction curve for larger PAHs.

\section{CONCLUSIONS}
\label{conclusion}
We presented UV-VIS spectroscopic studies of gas-phase condensed PAHs which have undergone a subsequent fractionation to produce mixtures of different size distributions. For the first time, we could demonstrate that MIS spectra of mixtures of large PAHs ($\geq$ 22 C atoms) can result in an almost smooth and featureless absorption in the visible and UV range. In contrast, PAH blends containing smaller molecules still produce narrow bands superimposed on a continuum in the UV. \citet{clayton03} have conducted a sensitive search for sharp UV absorption features in the diffuse ISM and failed to detect any which indicates the absence or low abundance, respectively, of free-flying small PAHs.

We also measured spectra of film-like deposits characterized by strong molecular interactions similar to those expected in PAH clusters bonded by van der Waals forces. In the ISM, the presence of such clusters has already been inferred from rather broad mid-IR emission features with a strong continuum \citep{Tielens08}. Clusters of PAH mixtures exhibit spectral similarities to nano-sized carbonaceous particles showing a pronounced UV bump. Its width and shape for PAH films containing more than 22 C atoms are nearly comparable to the interstellar extinction curve. However, its position is mainly determined by the mean size of the comprising PAHs. Based on calculations, we expect a redshift of the experimental bump position for larger PAHs and we propose that a distribution of molecules with a mean size of 50--60 carbon atoms can produce a bump close to 217.5 nm. This would be consistent with the expected size of interstellar PAHs obtained from the mid-IR emission bands. Small PAH molecules, which are formed in large quantities in condensation processes in the laboratory and, probably, in astrophysical environments, can be easily destroyed by the interstellar radiation field as concluded by several authors \citep{Jochims94, Allain96}.

Typical values for the carbon locked up in free-flying PAHs or small PAH clusters range from 22 to 65 C atoms per million H nuclei \citep{Tielens08, joblin92}. However, it should be taken into account that these estimates give rather lower limits for the PAH fraction because the assumption is made that all the power absorbed in the UV is re-emitted in the mid-IR despite the fact that many PAHs show strong luminescence in the visible. Furthermore, sufficiently large PAH clusters (very small grains) would contribute to the UV hump, but not to the mid-IR emission. Besides, typical interstellar sight lines used for UV studies probe regions of the ISM with a small column of dust while IR observations usually probe dense cloud environments \citep{clayton03}. Therefore, the PAH population responsible for the mid-IR features could be different from the one giving rise to features in the UV-VIS. We estimated that roughly C/H = (90 $\pm$ 30)$\times$10$^{-6}$ is required in order to produce a bump strength comparable to the mean interstellar UV hump. Free-flying neutral or cationic PAHs, as well as loosely bound PAH clusters of different sizes contribute. Calculations performed by \citet{pestellini08} indicate that the charge of the PAHs (at least for larger molecules) has no influence on the bump position. This can be plausibly explained by the argument that the density of electronic states in the UV (around 217.5 nm) is very high and the corresponding transitions are much lesser affected than the first few electronic states in the visible.\\

This work was supported by the Deutsche Forschungsgemeinschaft (DFG). We gratefully acknowledge Professor Dr. Klaus M\"{u}llen for providing the HBC sample and Dr. Hans Joachim R\"{a}der for performing MALDI--TOF measurements.
\\

\end{document}